\documentclass[12pt]{article}

\usepackage{latexsym}

\usepackage{graphicx}

\begin{document}


\title{Completely integrable sector  in 5D Einstein-Maxwell
 gravity and derivation of the dipole black ring solutions}

\author{
     Stoytcho S. Yazadjiev \thanks{E-mail: yazad@phys.uni-sofia.bg}\\
{\footnotesize  Department of Theoretical Physics,
                Faculty of Physics, Sofia University,}\\
{\footnotesize  5 James Bourchier Boulevard, Sofia~1164, Bulgaria }\\
}

\date{}

\maketitle

\begin{abstract}
We consider 5D Einstein-Maxwell (EM) gravity in spacetimes with
three commuting Killing vectors: one timelike and two spacelike Killing vectors one of
them being hypersurface-orthogonal. Assuming a special ansatz for the Maxwell field
we show that the 2-dimensional reduced EM equations are completely integrable by deriving a
Lax-pair presentation. We also develop a solution generating method for explicit construction of exact EM
solutions with considered symmetries. We also derive explicitly a new rotating
six parametric 5D EM solution which includes the dipole black ring solution as a particular case.
\end{abstract}


\sloppy

\section{Introduction}

In recent years the higher dimensional gravity is attracting much interest. Apart from the fact
that the higher dimensional gravity is interesting in its own right, the increasing amount of works
devoted to the study of the higher dimensional spacetimes is inspired from the string theory and
the brane-world scenario with large extra dimensions.
The gravity in higher dimensions exhibits much richer dynamics and spectrum  of
solutions than in four dimensions. One of the most reliable routes for  better understanding of higher
dimensional gravity and the related topics are the exact solutions. For example, recently discovered
exact black rings solutions with unusual horizon topology \cite{ER1,ER2}, demonstrated explicitly
that the 5D Einstein gravity
exhibits unexpected features completely absent in four dimensions. It was shown in \cite{ER2} that both
the black hole and the the black ring can carry the same conserved charges, the mass and a single angular
momentum, and therefore there is no uniqueness theorem in five dimensions.
Moreover, the black rings can also carry nonconserved charges which can be varied continuously without
altering the conserved charges. This fact leads to
continuous (classical) non-uniqness \cite{EMP}.

The higher dimensional solutions  found so far are not so many. As yet to the best of our knowledge
there are no EM solutions found in the literature that describe rotating charged black holes in
higher dimensions. However, some numerical solutions were recently  constructed in \cite{Kunz1}(see also \cite{Kunz2}).
Moreover the systematic construction
of new solutions in higher dimensions has not been accomplished in comparison with 4D case.
It is well known that both vacuum and electrovacuum 4D Einstein equations  are completely integrable
being restricted to spacetimes with two-dimensional Abelian group of isometries \cite{RG}-\cite{N}. This nice property
is also shared  by some effective string gravity models (or certain sectors of them) which allows us
to find many families of physically interesting exact solutions \cite{B}-\cite{YU}.
The $D$-dimensional vacuum
Einstein equations with $(D-2)$-dimensional Abelian group of isometries are completely integrable, too  \cite{DM2},\cite{POM}.

The aim of this work is to make a step in systematic construction of exact solutions in 5D
EM gravity. We show here that a certain sector of EM  gravity is completely integrable.
We also present an explicit method for generating exact 5D EM solutions from known solutions of the 5D vacuum Einstein equations.
As an illustration of the method we derive explicitly a new rotating
six parametric 5D EM solution which includes the dipole black ring solution as a particular case.

\section{Dimensional reduction, coset presentation and  complete integrability }

The 5D EM gravity is described by the field equations

\begin{eqnarray}\label{EMFE}
&&R_{\mu\nu} = {1\over 2} \left(F_{\mu\lambda}F_{\nu}^{\,\lambda}
 - {1\over 6} F_{\sigma\lambda}F^{\sigma\lambda} g_{\mu\nu}\right),  \\
&&\nabla_{\mu} F^{\mu\nu} = 0  \nonumber.
\end{eqnarray}

In this paper we consider 5D EM  gravity in spacetimes with  three commuting Killing vectors:
one timelike Killing vector $T$ and two spacelike Killing vectors $K_{1}$ and $K_{2}$.  We also assume
that the Killing vector $K_{2}$ is hypersurface orthogonal.

In adapted coordinates in which $K_{2}=\partial/\partial Y$, the spacetime
metric can be written into the form

\begin{equation}
ds^2 = e^{2u}dY^2 + e^{-u} h_{ij}dx^idx^j
\end{equation}

where $h_{ij}$ is a $4$-dimensional metric with Lorentz signature. Both $u$ and $h_{ij}$
depend on the coordinates $x^i$ only. The electromagnetic field is taken in the form\footnote{Throughout this paper
we denote the Killing vectors and their naturally corresponding 1-forms by the same letter. }

\begin{equation}
F = dA_{Y}\wedge dY.
\end{equation}

After a dimensional reduction along the Killing vector $K_{2}$, the field equations (\ref{EMFE})
are reduced to the following effective 4D theory:

\begin{eqnarray}
&&{\cal D}_{i}{\cal D}^{i}u =
- {1\over 3} e^{-2u}h^{ij}{\cal D}_{i}A_{Y} {\cal D}_{j}A_{Y},\\
&&{\cal D}_{i}\left(e^{- 2u}{\cal D}^{i}A_{Y} \right) = 0, \\
&&R(h)_{ij}= {3\over 2}\partial_{i}u\partial_{j}u
+ {1\over 2}e^{-2u}\partial_{i}A_{Y}\partial_{j}A_{Y}.
\end{eqnarray}

Here ${\cal D}_{i}$ and $R(h)_{ij}$ are the covariant derivative  and Ricci tensor with respect
to the  Lorentz metric $h_{ij}$. Let us introduce the symmetric matrix $M_{1}$ given by

\begin{eqnarray}
M_{1} = \left(%
\begin{array}{cc}
  e^{u} + {1\over 3}e^{-u}A^2_{Y} & {1\over \sqrt{3}} e^{-u}A_{Y} \\
 {1\over \sqrt{3}} e^{-u}A_{Y} & e^{-u} \\\end{array}%
\right)
\end{eqnarray}

with $\det M_{1}=1$. Then the dimensionally reduced EM equations become

\begin{eqnarray}
&&{\cal D}_{i}\left[{\cal D}^{i}M_{1}M^{-1}_{1}\right]=0 ,\\
&&R_{ij}(h) =  -{3\over 4} Tr\left[\partial_{i}M_{1}\partial_{j}M_{1}^{-1}\right].
\end{eqnarray}

These equations are yielded by the action

\begin{eqnarray}
S = {1\over 16\pi} \int d^4x \sqrt{h} \left[R(h) + {3\over 4}h^{ij} Tr\left(\partial_{i}M_{1}\partial_{j}M_{1}^{-1}\right) \right].
\end{eqnarray}

Clearly the action is invariant under the $SL(2,R)$ group where the group action is given by

\begin{eqnarray}
M_{1} \to GM_{1}G^{T},
\end{eqnarray}

$G \in SL(2,R)$. In fact the matrices $M_{1}$ parameterize a $SL(2,R)/SO(2)$ coset. So we obtain non-linear
$\sigma$-model coupled to 4D Einstein  gravity.

Next step is to further reduce the effective $4D$ theory along the Killing vectors $T$ and $K_{1}$.
In this connection it is useful to introduce the twist $\omega$ of the Killing vector $T$ defined by

\begin{eqnarray}\label{TD}
\omega = {1\over 2} \star (h)\left(T\wedge dT \right)
\end{eqnarray}

were $\star(h)$ is the Hodge dual with respect to the metric $h_{ij}$.

One can show that the Ricci 1-form ${\Re}_{h}[T]$  defined by
\begin{equation}
{\Re}_{h}[T] = R_{ij}(h)T^{j}dx^{i} ,
\end{equation}

 satisfies

\begin{equation}
\star(h)\left( T\wedge {\Re}_{h}[T] \right) = d\omega .
\end{equation}

Obviously, in our case we have ${\Re}_{h}[T]=0$, i.e. $d\omega$=0.  Therefore there exists (locally) a  potential $f$ such that

\begin{equation}\label{EFORM}
\omega = df.
\end{equation}

In adapted coordinates for the Killing vectors $T=\partial/\partial t$ and $K_{1}=\partial/\partial X$,
and in the canonical coordinates $\rho$ and $z$ for the transverse space, the 4D metric $h_{ij}$ can be written into the form

\begin{eqnarray}
h_{ij}dx^idx^j = -e^{2U}\left(dt + {\cal A} dX \right)^2 + e^{-2U}\rho^2 dX^2 + e^{-2U}e^{2\Gamma}(d\rho^2 + dz^2).
\end{eqnarray}

For this form of the metric $h_{ij}$, combining (\ref{TD}) and (\ref{EFORM}), and  after some algebra we find
that  the twist potential $f$ satisfies

\begin{eqnarray}\label{TPS}
\partial_{\rho}f &=& -{1\over 2} {e^{4U}\over \rho} \partial_{z}{\cal A} ,\\
\partial_{z} f &=& {1\over 2} {e^{4U}\over \rho} \partial_{\rho}{\cal A}.
\end{eqnarray}

Before writing the 2D reduced equations  we shall introduce the symmetric matrix

\begin{eqnarray}
M_{2} = \left(%
\begin{array}{cc}
  e^{2U} + 4f^2e^{-2U} & 2fe^{-2U} \\
 2fe^{-2U} & e^{-2U} \\\end{array}%
\right)
\end{eqnarray}

with $\det M_{2}=1$. Then the 2D reduced EM equations read

\begin{eqnarray}
&&\partial_{\rho}\left(\rho\partial_{\rho}M_{1} M^{-1}_{1} \right)
 + \partial_{z}\left(\rho\partial_{z}M_{1} M^{-1}_{1} \right) = 0 ,\\
&&\partial_{\rho}\left(\rho \partial_{\rho}M_{2}M^{-1}_{2} \right)
 + \partial_{z}\left(\rho \partial_{z}M_{2}M^{-1}_{2} \right) = 0 ,\\
\rho^{-1} \partial_{\rho}\Gamma &=&
- {1\over 8} \left[Tr\left(\partial_{\rho}M_{2}\partial_{\rho}M^{-1}_{2}\right)
 - Tr\left(\partial_{z}M_{2}\partial_{z}M^{-1}_{2}\right) \right]  \nonumber \\
&&- {3\over 8} \left[Tr\left(\partial_{\rho}M_{1}\partial_{\rho}M^{-1}_{1}\right)
- Tr\left(\partial_{z}M_{1}\partial_{z}M^{-1}_{1}\right) \right] ,\\
\rho^{-1} \partial_{z}\Gamma &=& - {1\over 4} Tr\left(\partial_{\rho}M_{2}\partial_{z}M^{-1}_{2}\right)
\nonumber \\
&& - {3\over 4} Tr\left(\partial_{\rho}M_{1}\partial_{z}M^{-1}_{1}\right).
\end{eqnarray}

As a result we find that that the "field variables" $M_{1}$ and $M_{2}$ satisfy the equations of two
$SL(2,R)/SO(2)$ $\sigma$-models in two dimensions, modified by the presence of the factor $\rho$.
The system equations for $\Gamma$
can be integrated, once a pair of solutions for the two $\sigma$-models is known. Therefore,
the problem of generating solutions to the 5D EM equations with the described symmetries reduces to
the solutions of the two $\sigma$-models.

It is well-known  that the $\sigma$-model equations are completely integrable \cite{BZ1,BZ2}. This is a consequence
of the fact that the $\sigma$-model equations can be considered as the compatibility condition of
the linear differential equations (Lax-pair presentation)\cite{BZ1,BZ2}

\begin{eqnarray}\label{LPP}
 D_{\rho} \Psi &=& {\rho {\cal U}+ \lambda V\over \lambda^2 + \rho^2} \Psi ,\\
 D_{z} \Psi &=& {\rho V - \lambda {\cal U}\over \lambda^2 + \rho^2} \Psi  \nonumber ,
\end{eqnarray}

where

\begin{eqnarray}
D_{\rho} = \partial_{\rho}  + {2\lambda\rho \over \lambda^2 + \rho^2}\partial_{\lambda}, \,\,\,\,
D_{z} = \partial_{z}  - {2\lambda^2 \over \lambda^2 + \rho^2}\partial_{\lambda}.
\end{eqnarray}

Here $V=\rho\partial_{z}M M^{-1}$, ${\cal U}=\rho\partial_{\rho}M M^{-1}$ and $\lambda$ is the complex
spectral parameter. The "wave function"  $\Psi(\rho,z,\lambda)$ is
a complex matrix. The $\sigma$-model equations then follows from the compatibility condition

\begin{equation}
[D_{\rho}, D_{z}]\Psi = 0.
\end{equation}

The matrix $M$ can be found from the "wave function" $\Psi$ as $M(\rho,z)=\Psi(\rho,z,\lambda=0)$.

The inverse scattering transform (IST) method can be directly applied to (\ref{LPP}) to generate multisoliton
solutions. The dressing procedure allows us to generate new solutions from known ones. Since this dressing
technique is well known we will not discuss it here and refer the reader to \cite{BZ1,BZ2}.

In this paper we will not apply the IST method. In the next section we  present new and simple enough
solution generating method which allows us to generate new 5D EM solutions from known solutions of the
5D vacuum  Einstein equations.

\section{Solution construction}

Let us consider two solutions $M_{1}=M^{(1)}$ and $M_{2}=M^{(2)}$ of the $\sigma$-model equations

\begin{equation}
\partial_{\rho}\left(\rho\partial_{\rho}M M^{-1} \right)
+ \partial_{z}\left(\rho\partial_{z}M M^{-1} \right) = 0 .\\
\end{equation}

In addition let us denote by $\gamma^{(i)}$ the solution of the system

\begin{eqnarray}
\rho^{-1} \partial_{z}\gamma^{(i)} &=&
-{1\over 4} Tr\left(\partial_{\rho}M^{(i)}\partial_{z}{M^{(i)}}^{-1} \right), \\
\rho^{-1} \partial_{\rho}\gamma^{(i)} &=&
-{1\over 8} \left[Tr\left(\partial_{\rho}M^{(i)}\partial_{\rho}{M^{(i)}}^{-1} \right)
- Tr\left(\partial_{z}M^{(i)}\partial_{z}{M^{(i)}}^{-1} \right) \right].
\end{eqnarray}

Then we find for the metric function $\Gamma$

\begin{equation}\label{GGGM}
\Gamma = \gamma^{(2)} + 3\gamma^{(1)}.
\end{equation}

From a practical point of view it is more convenient  to associate the $\sigma$-model solutions
$M^{(i)}$  with solutions of the vacuum Einstein equations\footnote{From now on all quantities
with subscript or superscript "E" correspond to the vacuum case.}

\begin{eqnarray}
ds^2_{E(i)} = e^{2u^{(i)}_{E}}dY^2 + e^{-u^{(i)}_{E}}\left [-e^{2{U^{(i)}_{E}}}\left(dt + {\cal A}^{(i)}_{E} dX \right)^2
\right. \\ \left. +  e^{-2{U^{(i)}_{E}}}\rho^2 dX^2 + e^{-2{U^{(i)}_{E}}}e^{2\Gamma^{(i)}_{E}}(d\rho^2 + dz^2)\right]\nonumber ,
\end{eqnarray}

which correspond to the matrixes

\begin{eqnarray}
M^{(i)} = \left(%
\begin{array}{cc}
  e^{2U^{(i)}_{E}} + 4\left(f^{(i)}_{E}\right)^2e^{-2U^{(i)}_{E}} & 2f^{(i)}_{E}e^{-2U^{(i)}_{E}} \\
 2f^{(i)}_{E}e^{-2U^{(i)}_{E}} & e^{-2U^{(i)}_{E}} \\\end{array}%
\right) .
\end{eqnarray}

The metric function $\Gamma^{(i)}_{E}$ for the vacuum Einstein equations can be found from the  equations
of $\Gamma$ by setting $A_{Y}=0$ in the matrix $M_{1}$. So we obtain

\begin{equation}\label{GGGMO}
\Gamma^{(i)}_{E} = \gamma^{(i)} + \Omega^{(i)}_{E}
\end{equation}

where $\Omega^{(i)}_{E}$ is a solution to the system

\begin{eqnarray}\label{OS}
\rho^{-1}\partial_{\rho}\Omega^{(i)}_{E} &=& {3\over 4}\left[\left(\partial_{\rho} u^{(i)}_{E}\right)^2
- \left(\partial_{z} u^{(i)}_{E}\right)^2 \right],\\
\rho^{-1}\partial_{z}\Omega^{(i)}_{E} &=& {3\over 2} \partial_{\rho} u^{(i)}_{E}\partial_{z} u^{(i)}_{E}.
\end{eqnarray}

We  then find from (\ref{GGGM}) and (\ref{GGGMO}) that

\begin{equation}
\Gamma = \Gamma^{(2)}_{E} - \Omega^{(2)}_{E} + 3\left[\Gamma^{(1)}_{E} - \Omega^{(1)}_{E} \right].
\end{equation}

Comparing the matrixes $M_{1}$ and $M^{(1)}$ we obtain

\begin{eqnarray}
e^{2u} = e^{4U^{(1)}_{E}} ,\\
A_{Y} = 2\sqrt{3} f^{(1)}_{E},
\end{eqnarray}

where $f^{(i)}_{E}$ satisfies\footnote{Clearly, these equations are restriction of (\ref{TPS}) to the vacuum case. }

\begin{eqnarray}\label{TPS}
\partial_{\rho}f^{(i)}_{E} &=& -{1\over 2} {e^{4U^{(i)}_{E}}\over \rho} \partial_{z}{\cal A}^{(i)}_{E} ,\\
\partial_{z} f^{(i)}_{E} &=& {1\over 2} {e^{4U^{(i)}_{E}}\over \rho} \partial_{\rho}{\cal A}^{(i)}_{E}.
\end{eqnarray}

Once having the metric function  $e^{2u}=g_{YY}$ we can write the EM metric

\begin{eqnarray}
ds^2 = e^{4U^{(1)}_{E}} dY^2 + e^{-2U^{(1)}_{E}} \left[ -e^{2U^{(2)}_{E}}\left(dt + {\cal A}^{(2)}_{E}dX \right)^2
+ e^{-2U^{(2)}_{E}}\rho^2 dX^2 \right. \nonumber \\
+\left.  \left(e^{2\Gamma^{(1)}_{E}} \over e^{2\Omega^{(1)}_{E} + {2\over 3}\Omega^{(2)}_{E} }  \right)^3
e^{-2U^{(2)}_{E}} e^{2\Gamma^{(2)}_{E}} (d\rho^2 + dz^2) \right].
\end{eqnarray}

Taking into account that

\begin{eqnarray}
g^{E(i)}_{00} &=& - e^{-u^{(i)}_{E}}e^{2U^{(i)}_{E}},\\
{\tilde g}^{E(i)}_{XX} &=& g^{E(i)}_{XX} - g^{E(i)}_{00}({\cal A}^{(i)}_{E})^2 = e^{-u^{(i)}_{E}}e^{-2U^{(i)}_{E}}\rho^2,\\
g^{E(i)}_{\rho\rho} &=& e^{-u^{(i)}_{E}}e^{-2U^{(i)}_{E}} e^{2\Gamma^{(i)}_{E}}  ,
\end{eqnarray}

and

\begin{eqnarray}
e^{4U^{(i)}_{E}} &=& (g^{E(i)}_{00})^2 g^{E(i)}_{YY},\\
e^{2\Gamma^{(i)}_{E}} &=& |g^{E(i)}_{00}|g^{E(i)}_{YY} g^{E(i)}_{\rho\rho},
\end{eqnarray}

the metric can be presented in more elegant form

\begin{eqnarray}
ds^2 &=& \left[|g^{E(1)}_{00}|\sqrt{g^{E(1)}_{YY}} \right]^2 dY^2
+ \left[\sqrt{g^{E(2)}_{YY}} \over |g^{E(1)}_{00}|\sqrt{g^{E(1)}_{YY}} \right]
\left[g^{E(2)}_{00}\left(dt + {\cal A}^{(2)}_{E}dX \right)^{2} +
{\tilde g}^{E(2)}_{XX}dX^2  \nonumber \right. \\ && \left. +
\left(|g^{E(1)}_{00}|g^{E(1)}_{YY} g^{E(1)}_{\rho\rho}
\over e^{2\Omega^{(1)}_{E}
+ {2\over 3}\Omega^{(2)}_{E}} \right)^3 g^{E(2)}_{\rho\rho} (d\rho^2 + dz^2) \right] .
\end{eqnarray}

Summarizing, we obtain  the following important result formulated as a
proposition.

{\bf Proposition.} {\it Let us consider two solutions of the vacuum
5D Einstein equations }

\begin{eqnarray}
ds_{E(i)}^2 = g^{E(i)}_{YY} dY^2 + g^{E(i)}_{00}\left(dt + {\cal A}^{(i)}_{E}dX \right)^{2} +
{\tilde g}^{E(i)}_{XX}dX^2 + g^{E(i)}_{\rho\rho} (d\rho^2 + dz^2)
\end{eqnarray}

{\it Then the following give a solution to the 5D EM equations\footnote{More generally
we can take $A_{Y}=\pm 2\sqrt{3} f^{(1)}_{E} + const$ which is
obvious.} }

\begin{eqnarray}
ds^2 &=& \left[|g^{E(1)}_{00}|\sqrt{g^{E(1)}_{YY}} \right]^2 dY^2
+ \left[\sqrt{g^{E(2)}_{YY}} \over |g^{E(1)}_{00}|\sqrt{g^{E(1)}_{YY}} \right]
\left[g^{E(2)}_{00}\left(dt + {\cal A}^{(2)}_{E}dX \right)^{2} +
{\tilde g}^{E(2)}_{XX}dX^2  \nonumber \right. \\ && \left. +
\left(|g^{E(1)}_{00}|g^{E(1)}_{YY} g^{E(1)}_{\rho\rho}
\over e^{2\Omega^{(1)}_{E}
+ {2\over 3}\Omega^{(2)}_{E}} \right)^3 g^{E(2)}_{\rho\rho} (d\rho^2 + dz^2) \right] ,\\
A_{Y} &=& 2\sqrt{3} f^{(1)}_{E} ,
\end{eqnarray}

{\it where  $f^{(1)}_{E}$ is a solution to the system  }

\begin{eqnarray}\label{TPS1}
\partial_{\rho}f^{(1)}_{E} &=& -{1\over 2} {(g^{E(1)}_{00})^2 g^{E(1)}_{YY}\over \rho} \partial_{z}{\cal A}^{(1)}_{E} ,\\
\partial_{z} f^{(1)}_{E} &=& {1\over 2} {(g^{E(1)}_{00})^2 g^{E(1)}_{YY}\over \rho} \partial_{\rho}{\cal A}^{(1)}_{E},
\end{eqnarray}

{\it and $\Omega^{(i)}_{E}$ satisfy }

\begin{eqnarray}\label{OS1}
\rho^{-1}\partial_{\rho}\Omega^{(i)}_{E} &=& {3\over 16}\left[\left(\partial_{\rho} \ln\left( g^{E(i)}_{YY}\right)\right)^2
- \left(\partial_{z} \ln \left(g^{E(i)}_{YY}\right)\right)^2 \right],\\
\rho^{-1}\partial_{z}\Omega^{(i)}_{E} &=&
{3\over 8} \partial_{\rho} \ln \left(g^{E(i)}_{YY}\right)\partial_{z}\ln\left( g^{E(i)}_{YY}\right).
\end{eqnarray}

Let us also note that, in general, the exchange of the two sigma models $M^{(1)} \longleftrightarrow M^{(2)}$
leads to different EM solutions.

The presented proposition gives us a tool to generate new 5D EM solutions in a simple way  from
known solutions to the vacuum 5D Einstein equations. The technical difficulties are eventually
concentrating in finding of $\Omega_{E}$ from  (\ref{OS}) and $f_{E}$ from (\ref{TPS}).

Through the use of the proposition we can generate the "5D EM images" of all known solutions of the vacuum
5D Einstein equations with the symmetries we consider here.
It is not possible to  present explicitly here  the "EM images" of all
known vacuum Einstein solutions. We shall demonstrate the application of the proposition on the case of rotating
neutral black rings generating in this way a new rotating six parametric EM solution
which includes he EM rotating dipole black ring solution as a  particular case.

\section{ Derivation of the rotating dipole black ring\\ solution}

The rotating dipole black ring solutions in 5D Einstein-Maxwell-dilaton (EMd) gravity were
given in \cite{EMP} without any derivation. What is said in \cite{EMP} is that these solutions can be
obtained form generalized $C$-metric \cite{EMP1} by double Wick rotation and analytic continuation of parameters.
As far as we are aware there is no explicit derivation of the dipole black ring solutions.
Here we generate a new (six parametric) rotating EM solution
and as a byproduct we give an explicit derivation of the EM rotating dipole black ring solution\footnote{
This solution can be obtained form the EMd dipole black ring  solutions in the limit when the dilaton coupling
parameter is zero.}.

The first step we should make in order to derive the EM dipole black ring solution is to
chose two known solutions of the vacuum 5D Einstein equations and to present them in canonical coordinates.
As should be expected, we take two copies of the neutral black ring solution with different parameters:
the first solution is with parameters $\{\lambda_{1},\nu_{1},{\cal R}_{1}\}$ while the second is parameterized
by $\{\lambda_{2},\nu_{2},{\cal R}_{2}\}$. It should be also noted that in the case under consideration
the Killing vectors are denoted by

\begin{equation}
K_{1} = {\partial/\partial \psi} , \,\,\,\, K_{2} = {\partial/\partial \phi}.
\end{equation}

The neutral black ring solution has already been written  in canonical coordinates in \cite{HAR},
that is why we present here the final formulas:

\begin{eqnarray}
|g^{E(i)}_{00}| &=& {(1+\lambda_{i})(1-\nu_{i})R^{(i)}_{1} + (1-\lambda_{i})(1+\nu_{i})R^{(i)}_{2}
-2(\lambda_{i} - \nu_{i})R^{(i)}_{3}
- \lambda_{i}(1-\nu_{i}^2){\cal R}_{i}^2  \over (1+\lambda_{i})(1-\nu_{i})R^{(i)}_{1}
+ (1-\lambda_{i})(1+\nu_{i})R^{(i)}_{2}
-2(\lambda_{i} - \nu_{i})R^{(i)}_{3}
+ \lambda_{i}(1-\nu_{i}^2){\cal R}_{i}^2 } , \nonumber \\
g^{E(i)}_{\Phi\Phi} &=& {(R^{(i)}_{3}+z - {1\over 2}{\cal R}_{i}^2 )(R^{(i)}_{2} - z
+ {1\over 2}{\cal R}_{i}^2\nu_{i})
\over R^{(i)}_{1} - z - {1\over 2}{\cal R}_{i}^2\nu_{i} }  \nonumber \\ &=& {(R^{(i)}_{1} + R^{(i)}_{2}
+ \nu_{i}{\cal R}^2_{i})  (R^{(i)}_{1} - R^{(i)}_{3} + {1\over 2}(1+ \nu_{i}){\cal R}^2_{i}) (R^{(i)}_{2}
+ R^{(i)}_{3} - {1\over 2}(1 - \nu_{i}){\cal R}^2_{i})
\over {\cal R}^2_{i} ((1-\nu_{i})R^{(i)}_{1} - (1+\nu_{i})R^{(i)}_{2} -2\nu_{i} R^{(i)}_{3}) } \nonumber \\
 g^{E(i)}_{\rho\rho} &=&  [(1+\lambda_{i})(1-\nu_{i})R^{(i)}_{1} + (1-\lambda_{i})(1+\nu_{i})R^{(i)}_{2}
 -2(\lambda_{i} - \nu_{i})R^{(i)}_{3}
+ \lambda_{i}(1-\nu_{i}^2){\cal R}_{i}^2 ] \nonumber \\
&& \times {(1-\nu_{i})R^{(i)}_{1} + (1+\nu_{i})R^{(i)}_{2} + 2\nu_{i} R^{(i)}_{3}
\over  8(1-\nu_{i }^2)^2 R^{(i)}_{1}R^{(i)}_{2}R^{(i)}_{3}} , \\
{\cal A}^{(i)}_{E} &=& {-2 C(\nu_{i},\lambda_{i}) {\cal R}_{i} (1-\nu_{i})
[R^{(i)}_{3} -R^{(i)}_{1} + {1\over 2}{\cal R}_{i}^2 (1+\nu_{i})] \over
(1+\lambda_{i})(1-\nu_{i})R^{(i)}_{1} + (1-\lambda_{i})(1+\nu_{i})R^{(i)}_{2} -2(\lambda_{i} - \nu_{i})R^{(i)}_{3}
- \lambda_{i}(1-\nu_{i}^2){\cal R}_{i}^2 } \nonumber
\end{eqnarray}

where

\begin{eqnarray}
R^{(i)}_{1} =\sqrt{\rho^2 + (z + {\nu_{i}\over 2}{\cal R}_{i}^2)^2 } , \\
R^{(i)}_{2} =\sqrt{\rho^2 + (z - {\nu_{i}\over 2}{\cal R}_{i}^2)^2 }, \\
R^{(i)}_{3} = \sqrt{\rho^2 + (z - {1\over 2}{\cal R}_{i}^2)^2 },\\
C(\nu_{i},\lambda_{i}) =  \sqrt{\lambda_{i}(\lambda_{i} -\nu_{i}) {1+\lambda_{i}\over 1- \lambda_{i} }} .
\end{eqnarray}

The next step is to find the functions $\Omega^{(i)}_{E}$ and $f^{(1)}_{E}$. After straightforward but tedious
calculations we obtain

\begin{eqnarray}
e^{{8\over 3} \Omega_{E}^{(i)}} &=& { [(1-\nu_{i})R^{(i)}_{1} + (1+\nu_{i})R^{(i)}_{2}
+ 2\nu_{i} R^{(i)}_{3}]^2\over 8(1-\nu_{i}^2)^2R^{(i)}_{1}R^{(i)}_{2}R^{(i)}_{3} }
g^{E(i)}_{\Phi\Phi} ,\\
f^{(i)}_{E} &=& {(1-\nu_{i}) {\cal R}_{i} C(\nu_{i},\lambda_{i}) [R^{(i)}_{1} - R^{(i)}_{3} +
{1\over 2}(1+ \nu_{i}) {\cal R}_{i}^2 ] \over (1+\lambda_{i})(1-\nu_{i})R^{(i)}_{1} +
(1-\lambda_{i})(1+\nu_{i})R^{(i)}_{2} + 2(\nu_{i}-\lambda_{i})R^{(i)}_{3}
+ \lambda_{i}(1-\nu_{i}^2){\cal R}_{i}^2 } \nonumber .
\end{eqnarray}

Once having the functions $\Omega^{(i)}_{E}$ and $f^{(1)}_{E}$ in explicit form  we can immediately apply
the proposition and
we obtain explicitly  a new EM solution presented in the canonical coordinates. The found EM solution
depends on six parameters $\{\lambda_{i}, \nu_{i}, {\cal R}_{i}, i=1,2 \}$ and, obviously, the solution is
very complicated. The detailed study of this new solution needs a separate investigation which we postpone for future publication.
Here we will consider only the particular case when

\begin{eqnarray}
\nu_{1}=\nu_{2}=\nu ,\,\,\, {\cal R}_{1} = {\cal R}_{2}={\cal R}.
\end{eqnarray}

In this case we also have

\begin{equation}
\Omega^{(1)}_{E} = \Omega^{(2)}_{E} ,\,\,\, R^{(1)}_{a} = R^{(2)}_{a} , a= 1,2,3
\end{equation}

which considerably simplifies the solution.
Even in this particular case the
solution looks complicated in the canonical coordinates. That is why it is more convenient to
present the solution in coordinates where it takes  simpler form. Such coordinates are the so-called
$C$-metric coordinates given by

\begin{eqnarray}
\rho = {{\cal R}^2 \sqrt{-G(x)G(y)}\over (x-y)^2 } ,\,\,\,
z = {1\over 2} {{\cal R}^2(1-xy)(2+\nu x + \nu y )\over (x-y)^2 }
\end{eqnarray}

where

\begin{eqnarray}
G(x) = (1-x^2)(1+\nu x),\\
-1\le x \le 1,\,\,\,\,  y\le -1.
\end{eqnarray}

Performing this coordinate change we find

\begin{eqnarray}
ds^2 &=&
\left[{F_{\lambda_{1}}(y) \over  F_{\lambda_{1}}(x) }\right]^2  {{\cal R}^2 G(x)\over (x-y)^2} d\phi^2
+ \left[{F_{\lambda_{1}}(x) \over  F_{\lambda_{1}}(y) }\right]
\left[ - {F_{\lambda_{2}}(y) \over  F_{\lambda_{2}}(x)}
\left(dt  + C(\nu,\lambda_{2}){\cal R} {1 + y\over F_{\lambda_{2}}(y) }d\psi  \right)^2  \right. \nonumber \\
&& \left. - {{\cal R}^2 F_{\lambda_{2}}(x)\over (x-y)^2 } {G(y)\over F_{\lambda_{2}}(y) }d\psi^2
+ F^3_{\lambda_{1}}(y) {{\cal R}^2  F_{\lambda_{2}}(x) \over (x-y)^2} \left({dx^2\over G(x)}
- {dy^2\over G(y) }\right)\right] ,\\
A_{\phi} &=& \pm \sqrt{3} C(\nu,\lambda_{1}) {\cal R} {1+ x\over F_{\lambda_{1}}(x)} + const .
\end{eqnarray}

The metric can be rearranged into the form

\begin{eqnarray}
ds^2 &=& -{F_{\lambda_{2}}(y) \over  F_{\lambda_{2}}(x)} {F_{\lambda_{1}}(x) \over  F_{\lambda_{1}}(y) }
\left(dt  + C(\nu,\lambda_{2}){\cal R} {1 + y\over F_{\lambda_{2}}(y) }d\psi  \right)^2  \\
&+& \left[F_{\lambda_{1}}(x) F^2_{\lambda_{1}}(y)\right] {{\cal R}^2 F_{\lambda_{2}}(x)\over (x-y)^2 }
\left[- {G(x)\over F^3_{\lambda_{1}}(y) F_{\lambda_{2}}(y) } d\psi^2 +  {dx^2\over G(x)}
- {dy^2\over G(y) }  +  {G(x)\over F^3_{\lambda_{1}}(x) F_{\lambda_{2}}(x) } d\phi^2 \right] \nonumber .
\end{eqnarray}

Finally, in order to exclude pathological behaviors of the metric  we must consider only negative $\lambda_{1}$, i.e.

\begin{equation}
\lambda_{1} = - \mu \,\,\, , 0\le\mu<1 .
\end{equation}

and positive $\lambda_{2}$ and $\nu$ satisfying

\begin{equation}
0<\nu \le \lambda_{2} <1.
\end{equation}

One can easily see that the generated 5D EM solution is just the EM rotating dipole black ring solution.
Let us also recall\cite{EMP} that in order to avoid the possible conical singularities at $x=\pm 1$ and $y=-1$
we must impose

\begin{eqnarray}
&&\Delta \phi = \Delta \psi = 2\pi {(1 + \mu)^{3/2} \sqrt{1-\lambda_{2}}\over 1-\nu } ,\\
&&{1-\lambda_{2}\over 1+\lambda_{2} } \left( {1+ \mu\over 1-\mu } \right)^3 = \left({1-\nu\over 1+\nu } \right)^2 .
\end{eqnarray}

\section{Conclusion}

In this paper we considered EM gravity in spacetimes admitting three commuting Killing vectors: one timelike and
two spacelike one of them being hypersurface orthogonal. Assuming also a special ansatz for the electromagnetic
field we have shown that  the EM equations reduce to two $SL(2,R)/SO(2)$ $\sigma$-models and
a separated linear system of first order partial differential equations. This ensures the existence of Lax-pair
presentation, therefore the complete integrability of the considered sector of EM gravity.
The Lax pair presentation also opens the way to apply the IST method and  to generate multisoliton
solutions.

Using the two $\sigma$-models structure of the reduced EM sector we gave an explicit construction
for generating exact 5D EM solutions from known solutions of the 5D vacuum Einstein equations in the same
symmetry sector. As an explicit example we constructed a six parameter rotating EM solution including
as a particular case the rotating EM dipole black ring solution. In this way we gave, for the first time,
an explicit derivation of the dipole black ring solution.

We shall conclude with some prospects for future work. Here we have shown that the "superposition"
of two neutral black ring solutions with certain parameters yields the dipole EM black ring solution
which schematically can be expressed as

\begin{equation}
\{neutral\,\, black\,\, ring\} + \{neutral\,\, black \,\,ring\} \to \{EM \,\,dipole \,\, black\,\, ring\}.
\end{equation}

It would be interesting to find the EM solutions corresponding to the schemes

\begin{eqnarray}
\{neutral\,\,  black\,\, hole\} &+& \{neutral\,\,  black\,\, hole\} \to \{ ?\}, \\
\{neutral\,\, black\,\, hole\}  &+& \{neutral\,\, black \,\,ring\} \to \{ ?\}, \\
\{neutral\,\, black \,\,ring\}  &+& \{neutral\,\, black\,\, hole\} \to  \{?\},
\end{eqnarray}

as well as other solutions. Some solutions of vacuum 5D Einstein equations which could serve as seeds
for new EM solutions are given in \cite{HAR}-\cite{AK}.

It would also be of interest to generalized this work for EM gravity in spacetimes with number of dimensions
greater than five and in the presence of a dilaton field non-minimaly coupled to
the electromagnetic field. Some results in these directions have already been found \cite{Y1}. They will be presented
elsewhere.

\section*{Acknowledgements}
I would like to thank I. Stefanov for reading the manuscript.
This work was partially supported by the
Bulgarian National Science Fund under Grant MUF04/05 (MU 408)
and the Sofia University Research Fund.

\end{document}